\documentclass[12pt]{article}
\usepackage{amsfonts}

\usepackage{graphicx}
\usepackage{amsmath}
\usepackage{float}

\textwidth=6.5in \hoffset=-.75in \voffset=-1in \numberwithin{equation}{section}
\numberwithin{figure}{section}

\input{tcilatex}

\begin{document}

\begin{titlepage}
\bigskip \begin{flushright}
hep-th/0408189\\
\end{flushright}
\vspace{1cm}
\begin{center}
{\Large \bf {Atiyah-Hitchin M-Branes}}\\
\end{center}
\vspace{2cm}
\begin{center}
 A. M.
Ghezelbash{ \footnote{ EMail: amasoud@avatar.uwaterloo.ca}}, R. B. Mann{
\footnote{ EMail: mann@avatar.uwaterloo.ca}}\\
Department of Physics, University of Waterloo, \\
Waterloo, Ontario N2L 3G1, Canada\\
\vspace{1cm}
\end{center}

\begin{abstract}
We present new M2 and M5 brane solutions in M-theory based on transverse 
Atiyah-Hitchin space and other self-dual geometries. One novel feature of these 
solutions is that they have bolt-like fixed points yet still preserve 1/4 of the 
supersymmetry. All the solutions can be reduced down to ten dimensional 
intersecting  brane configurations. 
\end{abstract}
\end{titlepage}\onecolumn 
\bigskip 

\section{Introduction}

Fundamental M-theory in the low-energy limit is generally believed to be
effectively described by $D=11$ supergravity \cite{gr1,gr2,gr3}. This
suggests that brane solutions in the latter theory furnish classical soliton
states of M-theory, motivating considerable interest in this subject. There
is particular interest in supersymmetric $p$-brane solutions that saturate
the BPS bound upon reduction to 10 dimensions. Some supersymmetric solutions
of two or three orthogonally intersecting 2-branes and 5-branes in $D=11$\
supergravity were obtained some years ago \cite{Tsey}, and more such
solutions have since been found \cite{other}.

Recently interesting new supergravity solutions for localized D2/D6, D2/D4,
NS5/D6 and NS5/D5 intersecting brane systems were obtained \cite%
{hashi,CGMM2,CGMM5}. By lifting a D6 (D5 or D4)-brane to four-dimensional
Taub-NUT/Bolt and Eguchi-Hanson geometries embedded in M-theory, these
solutions were constructed by placing M2- and M5-branes in the Taub-NUT/Bolt
and Eguchi-Hanson background geometries. The special feature of these
constructions\ is that the solution is not restricted to be in the near core
region of the D6 (D5 or D4)-brane.\ \ 

Taub-NUT space is a special case of the Atiyah-Hitchin space and since the
building blocks of M-theory are M2- and M5-branes, it is natural to
investigate the possibility of placing M2- and M5-branes in the
Atiyah-Hitchin background space. This is the subject of the present paper,
in which we consider the embedding of\ Atiyah-Hitchin geometry in M-theory
with an M2- or M5-brane.\ For all of the different solutions we obtain, 1/4
of the supersymmetry is preserved, and the metric has bolt-like fixed points
(i.e. of maximal co-dimensionality). This is an interesting feature of all
the solutions we obtain, quite distinct from all previously constructed
M-brane solutions with bolt-like fixed points \cite{CGMM2,CGMM5}, for which
no supersymmetries are preserved. The difference arises as a result of the
self-duality of the Atiyah-Hitchin metric compared to non-self-dual
Taub-Bolt metrics. In the former case, self-duality preserves some
supersymmetry while in the latter case, the lack of self-duality precludes
any possible supersymmetry. We then compactify these solutions on a circle,
obtaining the different fields of type IIA string theory. Explicit
calculation shows that in all cases the metric is asymptotically (locally)
flat, though for some of our compactified solutions the type IIA dilaton
field diverges at infinity.

The Atiyah-Hitchin space is a part of the set of two monopole solutions of
Bogomol'nyi equation. The moduli space of solutions is of the form%
\begin{equation*}
\mathbb{R}^{3}\otimes \frac{S^{1}\otimes \mathcal{M}}{\mathbb{Z}_{2}}
\end{equation*}%
where the factor $\mathbb{R}^{3}\otimes S^{1}$ describes the center of mass
of two monopoles and a phase factor that is related to the total electric
charge of the system. The interesting part of the moduli space is the four
dimensional manifold $\mathcal{M}$, which has self-dual curvature. The
self-duality comes from the hyper-K\"{a}hler property of the moduli space.
Since $\mathbb{R}^{3}\otimes S^{1}$ is flat and decouples from $\mathcal{M}$%
, the four dimensional manifold $\mathcal{M}$ should be hyper-K\"{a}hler,
which is equivalent to a metric with self-dual curvature in four dimensions.
The manifold $\mathcal{M}$ describes the separation of the two monopoles and
their relative phase angle (or electric charges). A further aspect
concerning $\mathcal{M}$ is that it should be $SO(3)$\ invariant, since two
monopoles do exist in ordinary flat space; hence the metric on $\mathcal{M}$
can be expressed in terms of three functions of the monopole separation.
Self-duality implies that these three functions obey a set of first-order
ordinary differential equations.

In recent years, this space and its various generalizations were identified
with the full quantum moduli space of $\mathcal{N}=4$ supersymmetric gauge
theories in three dimensions \cite{seib}.

The outline of our paper is as follows. In section \ref{sec:review}, we
discuss briefly the field equations of supergravity, the M2- and M5-brane
metrics and the Killing spinor equations. In section \ref{sec:M2}, we
present the different M2-brane solutions that preserve 1/4 of the
supersymmetry. We find type IIA D2$\perp $D6(2) intersecting brane solutions
upon dimensional reduction. \ In section \ref{sec:sc} the alternative
M2-brane solutions are presented. These solutions are obtained by
continuation of the real separation constant into a pure imaginary
separation constant. In section \ref{sec:M5}, we present different M5
solutions that preserve 1/4 of the supersymmetry; upon dimensional reduction
at infinity, we find IIA NS5$\perp $D6(5) intersecting brane solutions.

\section{M2- and M5- Branes and Kaluza-Klein Reduction}

\label{sec:review}

The equations of motion for eleven dimensional supergravity when we have
maximal symmetry (i.e. for which the expectation values of the fermion
fields is zero), are \cite{DuffKK}%
\begin{eqnarray}
R_{mn}-\frac{1}{2}g_{mn}R &=&\frac{1}{3}\left[ F_{mpqr}F_{n}^{\phantom{n}%
pqr}-\frac{1}{8}g_{mn}F_{pqrs}F^{pqrs}\right]  \label{GminGG} \\
\nabla _{m}F^{mnpq} &=&-\frac{1}{576}\varepsilon ^{m_{1}\ldots
m_{8}npq}F_{m_{1}\ldots m_{4}}F_{m_{5}\ldots m_{8}}  \label{dF}
\end{eqnarray}%
where the indices $m,n,\ldots $ are 11-dimensional world space indices. For
an M2-brane, we use the metric and four-form field strength 
\begin{equation}
ds_{11}^{2}=H(y,r)^{-2/3}\left( -dt^{2}+dx_{1}^{2}+dx_{2}^{2}\right)
+H(y,r)^{1/3}\left( d\mathfrak{s}_{4}^{2}(y)+ds_{4}^{2}(r)\right)
\label{ds11genM2}
\end{equation}%
and non-vanishing four-form field components%
\begin{equation}
F_{tx_{1}x_{2}y}=-\frac{1}{2H^{2}}\frac{\partial H}{\partial y}~\
~,~~~F_{tx_{1}x_{2}r}=-\frac{1}{2H^{2}}\frac{\partial H}{\partial r}.
\end{equation}%
and for an M5-brane, the metric and four-form field strength are

\begin{eqnarray}
ds^{2} &=&H(y,r)^{-1/3}\left( -dt^{2}+dx_{1}^{2}+\ldots +dx_{5}^{2}\right)
+H(y,r)^{2/3}\left( dy^{2}+ds_{4}^{2}(r)\right) ~~~  \label{ds11general} \\
F_{m_{1}\ldots m_{4}} &=&\frac{\alpha }{2}\epsilon _{m_{1}\ldots
m_{5}}\partial ^{m_{5}}H~\ ~,~~~\alpha =\pm 1  \label{Fgeneral}
\end{eqnarray}%
where $d\mathfrak{s}_{4}^{2}(y)$ and $ds_{4}^{2}(r)$ are two
four-dimensional (Euclideanized) metrics, depending on the non-compact
coordinates $y$ and $r$, respectively and the quantity $\alpha =\pm 1,$\
which corresponds to an M5 brane and an anti-M5 brane respectively. The
general solution, where the transverse coordinates are given by a flat
metric, admits a solution with 16 Killing spinors \cite{gauntlett}.

The 11D metric and four-form field strength can be easily reduced down to
ten dimensions using the following equations%
\begin{eqnarray}
g_{mn} &=&\left[ 
\begin{array}{cc}
e^{-2\Phi /3}\left( g_{\alpha \beta }+e^{2\Phi }C_{\alpha }C_{\beta }\right)
& \nu e^{4\Phi /3}C_{\alpha } \\ 
\nu e^{4\Phi /3}C_{\beta } & \nu ^{2}e^{4\Phi /3}%
\end{array}%
\right]  \label{FKKreduced1} \\
F_{(4)} &=&\mathcal{F}_{(4)}+\mathcal{H}_{(3)}\wedge dx_{10}
\label{FKKreduced2}
\end{eqnarray}

Here $\nu $ is the winding number (the number of times the M5 brane wraps
around the compactified dimensions) and $x_{10}$ is the eleventh dimension,
on which we compactify. We use hats in the above to differentiate the
eleven-dimensional fields from the ten-dimensional ones that arise from
compactification. $\mathcal{F}_{(4)}$ and $\mathcal{H}_{(3)}$ are the RR
four-form and the NSNS three-form field strengths corresponding to $%
A_{\alpha \beta \gamma }$ and $B_{\alpha \beta }$.

The number of non-trivial solutions to the Killing spinor equation 
\begin{equation}
\partial _{m}\epsilon +\frac{1}{4}\omega _{abm}\Gamma ^{ab}\epsilon +\frac{1%
}{144}\Gamma _{m}^{\phantom{m}npqr}F_{npqr}\epsilon -\frac{1}{18}\Gamma
^{pqr}F_{mpqr}\epsilon =0  \label{killingspinoreq}
\end{equation}%
determine the amount of supersymmetry of the solution, where the $\omega $'s
are the spin connection coefficients, and $\Gamma ^{a_{1}\ldots
a_{n}}=\Gamma ^{\lbrack a_{1}}\ldots \Gamma ^{a_{n}]}$. The indices $a,b,...$
are 11 dimensional tangent space indices and the $\Gamma ^{a}$ matrices are
the eleven dimensional equivalents of the four dimensional Dirac gamma
matrices, and must satisfy the Clifford algebra 
\begin{equation}
\left\{ \Gamma ^{a},\Gamma ^{b}\right\} =-2\eta ^{ab}  \label{cliffalg}
\end{equation}

In ten dimensional type IIA string theory, we can have D-branes or
NS-branes. D$p$-branes can carry either electric or magnetic charge with
respect to the RR fields; the metric takes the form \cite{gauntlett}%
\begin{equation}
ds_{10}^{2}=f^{-1/2}\left( -dt^{2}+dx_{1}^{2}+\ldots +dx_{p}^{2}\right)
+f^{1/2}\left( dx_{p+1}^{2}+\ldots +dx_{9}^{2}\right)  \label{gDpbrane}
\end{equation}%
where the harmonic function\thinspace\ $f$ generally depends on the
transverse coordinates.

An NS5-brane carries a magnetic two-form charge; the corresponding metric
has the form 
\begin{equation}
ds_{10}^{2}=-dt^{2}+dx_{1}^{2}+\ldots +dx_{5}^{2}+f\left( dx_{6}^{2}+\ldots
+dx_{9}^{2}\right)  \label{gNS5brane}
\end{equation}%
In what follows we will obtain a mixture of D-branes and NS-branes.

\section{Embedding Atiyah-Hitchin space in an M2-brane metric}

\label{sec:M2}

The eleven dimensional M2-brane with an embedded transverse Atiyah-Hitchin
space is given by the following metric%
\begin{equation}
ds_{11}^{2}=H(y,r)^{-2/3}\left( -dt^{2}+dx_{1}^{2}+dx_{2}^{2}\right)
+H(y,r)^{1/3}\left( dy^{2}+y^{2}d\Omega _{3}^{2}+ds_{AH}^{2}\right)
\label{ds11m2}
\end{equation}%
and non-vanishing four-form field components%
\begin{equation}
F_{tx_{1}x_{2}y}=-\frac{1}{2H^{2}}\frac{\partial H}{\partial y}~\
~,~~~F_{tx_{1}x_{2}r}=-\frac{1}{2H^{2}}\frac{\partial H}{\partial r}.
\label{Fm2}
\end{equation}%
The Atiyah-Hitchin metric $ds_{AH}^{2}$ is given by the following manifestly 
$SO(3)$ invariant form \cite{GM} 
\begin{equation}
ds_{AH}^{2}=f^{2}(r)dr^{2}+a^{2}(r)\sigma _{1}^{2}+b^{2}(r)\sigma
_{2}^{2}+c^{2}(r)\sigma _{3}^{2}  \label{AHmetric}
\end{equation}%
with 
\begin{equation}
\begin{array}{c}
\sigma _{1}=-\sin \psi d\theta +\cos \psi \sin \theta d\phi \\ 
\sigma _{2}=\cos \psi d\theta +\sin \psi \sin \theta d\phi \\ 
\sigma _{3}=d\psi +\cos \theta d\phi%
\end{array}
\label{mcFORMS}
\end{equation}%
where $\sigma _{i\text{ }}$are Maurer-Cartan one-forms with the property 
\begin{equation}
d\sigma _{i}=\frac{1}{2}\varepsilon _{ijk}\sigma _{j}\wedge \sigma _{k}.
\label{dsigma}
\end{equation}%
We note that the metric on the $\mathbb{R}^{4}$ (with a radial coordinate $R$
and Euler angles ($\theta ,\phi ,\psi $) on an $S^{3}$) could be written in
terms of Maurer-Cartan one-forms by 
\begin{equation}
ds^{2}=dR^{2}+\frac{R^{2}}{4}(\sigma _{1}^{2}+\sigma _{2}^{2}+\sigma
_{3}^{2}).  \label{s3METRIC}
\end{equation}%
We also note that $\sigma _{1}^{2}+\sigma _{2}^{2}$ is the standard metric
of the round unit radius $S^{2}$ and $4(\sigma _{1}^{2}+\sigma
_{2}^{2}+\sigma _{3}^{2})$ gives the same for $S^{3}.$ The metric (\ref%
{AHmetric}) satisfies Einstein's equations provided%
\begin{equation}
\begin{array}{c}
a^{\prime }=f\frac{(b-c)^{2}-a^{2}}{2bc} \\ 
b^{\prime }=f\frac{(c-a)^{2}-b^{2}}{2ca} \\ 
c^{\prime }=f\frac{(a-b)^{2}-c^{2}}{2ab}.%
\end{array}
\label{conditions}
\end{equation}%
Choosing\textbf{\ }$f(r)=-\frac{b(r)}{r}$\textbf{\ }the explicit expressions
for the metric functions $a,b$ and $c$ are given by%
\begin{equation}
\begin{array}{c}
a(r)=\sqrt{\frac{r\Upsilon \sin (\gamma )\{\frac{1-\cos (\gamma )}{2}r-\sin
(\gamma )\Upsilon \}}{\Upsilon \sin (\gamma )+r\cos ^{2}(\frac{\gamma }{2})}}
\\ 
b(r)=\sqrt{\frac{\{\Upsilon \sin (\gamma )-\frac{1-\cos \gamma }{2}%
r\}r\{-\Upsilon \sin (\gamma )-\frac{1+\cos \gamma }{2}r\}}{\Upsilon \sin
(\gamma )}} \\ 
c(r)=-\sqrt{\frac{r\Upsilon \sin (\gamma )\{\frac{1+\cos (\gamma )}{2}r+\sin
(\gamma )\Upsilon \}}{-\Upsilon \sin (\gamma )+\frac{1-\cos \gamma }{2}r}}%
\end{array}
\label{abc}
\end{equation}%
where 
\begin{equation}
\Upsilon =\frac{2nE\{\sin (\frac{\gamma }{2})\}}{\sin (\gamma )}-\frac{%
nK\{\sin (\frac{\gamma }{2})\}\cos (\frac{\gamma }{2})}{\sin (\frac{\gamma }{%
2})}  \label{GAMMA}
\end{equation}%
and 
\begin{equation}
K(\sin (\frac{\gamma }{2}))=\frac{r}{2n}.  \label{gama}
\end{equation}%
In the above equations, $K$ and $E$\ are the elliptic integrals 
\begin{eqnarray}
K(k) &=&\int_{0}^{1}\frac{dt}{\sqrt{1-t^{2}}\sqrt{1-k^{2}t^{2}}}%
=\int_{0}^{\pi /2}\frac{d\theta }{\sqrt{1-k^{2}\cos ^{2}\theta }}
\label{Ell} \\
E(k) &=&\int_{0}^{1}\frac{\sqrt{1-k^{2}t^{2}}dt}{\sqrt{1-t^{2}}}%
=\int_{0}^{\pi /2}\sqrt{1-k^{2}\cos ^{2}\theta }d\theta
\end{eqnarray}%
and the coordinate $r$ ranges over the interval $[n\pi ,\infty )$, which
corresponds to $\gamma \in \lbrack 0,\pi ).$ The positive number $n$\ is a
constant number with unit of length that is related to NUT charge of metric
at infinity obtained from Atiyah-Hitchin metric.

In fact as $r\rightarrow \infty ,$ the metric (\ref{AHmetric}) reduces to 
\begin{equation}
ds_{AH}^{2}\rightarrow (1-\frac{2n}{r})(dr^{2}+r^{2}d\theta ^{2}+r^{2}\sin
^{2}\theta d\phi ^{2})+4n^{2}(1-\frac{2n}{r})^{-1}(d\psi +\cos \theta d\phi
)^{2}  \label{reducedAH}
\end{equation}%
which is the well known Euclidean Taub-NUT metric with a negative NUT charge 
$N=-n.$ One can compactify the M2-brane solution at infinity over the circle
described by the coordinate \ $\psi $, which from equation (\ref{reducedAH})
has radius $2n$.

\begin{figure}[tbp]
\centering                                                                
\begin{minipage}[c]{.3\textwidth}
        \centering
        \includegraphics[width=\textwidth]{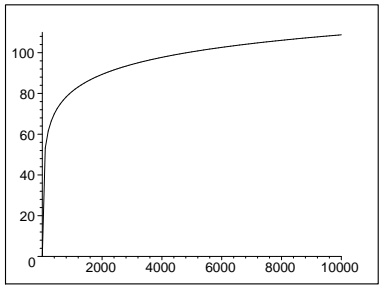}
    \end{minipage}
\caption{ Numerical solution of the radial equation (\ref{M2eqR}) for $%
R_{k}/10^{7}$ as a function of $\frac{1}{r-\protect\pi },$ where we set $n=1$
for simplicity. So for $r\approx \protect\pi $, $R$ diverges logarithmically
as $\ln (\frac{1}{r-\protect\pi })$. Note that the plot is only reliable for
large $\frac{1}{r-\protect\pi }$ .}
\label{M2Rnumeric}
\end{figure}

The above metric is obtained from a consideration of the limiting behaviors
of the functions $a,b$ and $c$ at large monopole separation that are given by%
\begin{equation}
\begin{array}{c}
a(r)=r(1-\frac{2n}{r})^{1/2}+O(e^{-r/n}) \\ 
b(r)=r(1-\frac{2n}{r})^{1/2}+O(e^{-r/n}) \\ 
c(r)=-2n(1-\frac{2n}{r})^{-1/2}+O(e^{-r/n}).%
\end{array}
\label{abcatinfinity}
\end{equation}%
The metric (\ref{ds11m2}) is a solution to the eleven dimensional
supergravity equations provided $H\left( y,r\right) $ is a solution to the
differential equation 
\begin{equation}
r^{2}\frac{\partial ^{2}H}{\partial r^{2}}+r\{r\frac{(ac)^{\prime }}{ac}+1\}%
\frac{\partial H}{\partial r}+b^{2}\{\frac{\partial ^{2}H}{\partial y^{2}}+%
\frac{3}{y}\frac{\partial H}{\partial y}\}=0.  \label{tndiffeq}
\end{equation}%
This equation is straightforwardly separable. Substituting 
\begin{equation}
H(y,r)=1+Q_{M2}Y(y)R(r)  \label{Hyrsep}
\end{equation}%
where $Q_{M2}$ is the charge on the M2 brane, we arrive at two differential
equations for $Y(y)$ and $R(r).$ The solution of the differential equation
for $Y(y)$ is 
\begin{equation}
Y(y)=\frac{J_{1}(ky)}{y}  \label{Y1}
\end{equation}%
which has a damped oscillating behavior at infinity. The differential
equation for $R(r)$ is

\begin{equation}
acr^{2}\frac{d^{2}R_{k}(r)}{dr^{2}}+\{acr-\frac{1}{2}%
r[(a-b)^{2}+(b-c)^{2}-a^{2}-c^{2}]\}\frac{dR_{k}(r)}{dr}%
-k^{2}ab^{2}cR_{k}(r)=0  \label{M2eqR}
\end{equation}%
where we have used the equations (\ref{conditions}) and $k$ is the
separation constant. Although equation (\ref{M2eqR}) does not have any
analytic closed solution, we can solve it numerically. A typical numerical
solution of (\ref{M2eqR}) is given in figure \ref{M2Rnumeric}, where for
simplicity we set $n=1$. The qualitative behaviour of the numerical
solutions of equation (\ref{M2eqR}) for other values of $n$\ is similar to
figure \ref{M2Rnumeric}, with the logarithmic divergence shifting to the
point $r\simeq \pi n.$

The most interesting point is near $r\simeq \pi n$. Indeed, the plot in
figure \ref{M2Rnumeric} is reliable only near $r\simeq \pi n,$\
corresponding to very large values on the horizontal axis. The divergence of
the radial function at $r\simeq \pi n$ is given by%
\begin{equation}
R_{k}(r)\simeq K_{0}(k(r-\pi n))=-\ln (\frac{k}{2})-\gamma -\ln (r-\pi
n)+O\{(r-\pi n)^{2}\}  \label{RatreqPi}
\end{equation}%
where $K_{0}$ is the modified Bessel function of the second kind and $\gamma 
$ is the Euler-Mascheroni constant. 
\ 

The behaviour of $\ a,b$ and $c$ near $r\simeq \pi n$ is%
\begin{equation}
\begin{array}{c}
a(r)=2(r-\pi n)+O((r-\pi n)^{2}) \\ 
b(r)=\pi n+\frac{1}{2}(r-\pi n)+O((r-\pi n)^{2}) \\ 
c(r)=-\pi n+\frac{1}{2}(r-\pi n)+O((r-\pi n)^{2})%
\end{array}
\label{abcofr}
\end{equation}%
which indicates a bolt singularity at this point since $a(r)\simeq 0$. \ By
using the $SO(3)$ invariance of the metric, we can write the metric element (%
\ref{AHmetric}) near the bolt location as%
\begin{equation}
ds^{2}=dr^{2}+4(r-\pi n)^{2}(d\widetilde{\psi }+\cos \widetilde{\theta }d%
\widetilde{\phi })^{2}+\pi ^{2}n^{2}(d\widetilde{\theta }+\sin ^{2}%
\widetilde{\theta }d\widetilde{\phi }).  \label{nearbolt}
\end{equation}%
where $\widetilde{\psi },\widetilde{\theta }$\ and $\widetilde{\phi }$\ are
a new set of Euler angles related to $\psi ,\theta ,$\ $\phi $\ by%
\begin{equation}
\mathcal{R}_{1}(\widetilde{\psi })\mathcal{R}_{3}(\widetilde{\theta })%
\mathcal{R}_{1}(\widetilde{\phi })=\mathcal{R}_{3}(\psi )\mathcal{R}%
_{2}(\theta )\mathcal{R}_{3}(\phi )  \label{Euler}
\end{equation}%
in which $\mathcal{R}_{i}(\alpha )$\ represents a rotation by $\alpha $\
about the $i$th axis. Note that the last term in (\ref{nearbolt}) is the
induced metric on the two dimensional bolt.

\begin{figure}[tbp]
\centering                                    
\begin{minipage}[c]{.3\textwidth}
        \centering
        \includegraphics[width=\textwidth]{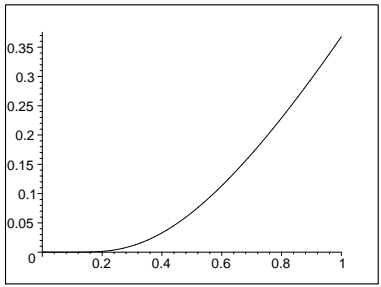}
    \end{minipage}
\caption{ Behavior of the radial function $R_{k}$ as a function of $\frac{1}{%
r},$ near infinity.}
\label{M2Ratlarger}
\end{figure}

To find the behaviour of the radial function $R_{k}(r)$ at large $r$, we use
the relations (\ref{abcatinfinity}); in this case we find%
\begin{equation}
R_{k}(r)\simeq \frac{e^{-\frac{k^{2}}{\left| k\right| }r}}{r}
\label{Ratlarger}
\end{equation}%
a plot of which is given in figure \ref{M2Ratlarger}.

The final solution will be a superposition of all possible solutions and
takes the form 
\begin{equation}
H(y,r)=1+Q_{M2}\int_{0}^{\infty }dkp(k)\frac{J_{1}(ky)}{y}R_{k}(r)
\label{gensolutionnearreqPi}
\end{equation}%
where $p(k)$ is the measure function.

As $r\rightarrow \infty $\ the Atiyah-Hitchin metric reduces to the $N=-n$\
Taub-NUT metric. It is tempting to use this to fix the form of the measure
function $p(k)$\ via comparison to the M2-brane solution obtained by
embedding a Taub-NUT space \cite{hashi, CGMM2}. \ However this is not
correct; the derivation of the measure function for the Taub-NUT based
M2-brane (with positive NUT charge $\mathcal{N}$) assumed that $r<<\mathcal{N%
}$\ \cite{hashi, CGMM2}. In the present case the Taub-NUT metric (\ref%
{reducedAH}) is well defined only for $r\geq 2n.$

To fix $p(k)$\ we compare the relation (\ref{gensolutionnearreqPi}) to that
of the metric function of an M2-brane in a transverse flat metric\ \ $%
\mathbb{R}^{4}\otimes \mathbb{R}^{2}\otimes S^{2}$, obtained by looking at
the near bolt limit. At $r\simeq \pi n,$\ for a fixed point on the bolt, the
metric reduces to the metric of $\mathbb{R}^{2}$, that is $d\widetilde{r}%
^{2}+4\widetilde{r}^{2}d\widetilde{\psi }^{2}$\ where $\widetilde{r}=r-n\pi
. $\ In this case the radial function is given by (\ref{RatreqPi}); to
obtain reduction of the metric function (\ref{gensolutionnearreqPi}) to $1+%
\frac{Q_{M2}}{\mathcal{R}^{6}}$\ where $\mathcal{R}=\sqrt{y^{2}+\widetilde{r}%
^{2}}$, we must fix the measure function to be $p(k)\propto k^{4}$.
Absorbing the constant into the M2-brane charge yields%
\begin{equation}
H_{AH}(y,r)=1+Q_{M2}\int_{0}^{\infty }dkk^{4}\frac{J_{1}(ky)}{y}R_{k}(r)
\label{AHgeneralH}
\end{equation}%
as the metric function of the M2-brane solution (\ref{ds11m2}).

Since $\partial /\partial \psi $\ is not a Killing vector (except for $%
a(r)=b(r)$) we cannot use the reduction relations (\ref{FKKreduced1}) and (%
\ref{FKKreduced2}) to find 10D type IIA fields explicitly. However, we note
that at large $r$, we have $a(r)=b(r)$\ up to terms of order of $e^{-r/n}$.
Hence the reduction relations\ (\ref{FKKreduced1}) and (\ref{FKKreduced2})
at large $r$\ yield the explicit fields in 10D that describe a D2$\perp $%
D6(2) system which preserves 1/4 of the supersymmetry \cite{CGMM2}. So
although we cannot explicitly write the 10D brane system fields
corresponding to the M2-brane solution (\ref{ds11m2}), the asymptotic
behaviour of the system is given by the known fields of D2$\perp $D6(2)
system which preserves 1/4 of the supersymmetry \cite{CGMM2}.

The second possible M2-brane solution is given by%
\begin{equation}
ds_{11}^{2}=H(y,r)^{-2/3}\left( -dt^{2}+dx_{1}^{2}+dx_{2}^{2}\right)
+H(y,r)^{1/3}\left( ds_{TN}^{2}+ds_{AH}^{2}\right)  \label{ds11m2withTN}
\end{equation}%
where 
\begin{eqnarray}
ds_{TN}^{2} &=&f(y)\left( dy^{2}+y^{2}(d\alpha ^{2}+\sin ^{2}(\alpha )d\beta
^{2})\right) +\left( \frac{(4\mathbf{n})^{2}}{f(y)}\right) \left( d\sigma +%
\frac{1}{2}\cos (\alpha )d\beta \right) ^{2}  \label{dstn4hashi} \\
f(y) &=&\left( 1+\frac{2\mathbf{n}}{y}\right) .
\end{eqnarray}%
In this case, after separation of variables by the relation (\ref{Hyrsep}),
we find the same differential equation for $R(r)$\ as given in equation (\ref%
{M2eqR}), and the solution of the differential equation for $Y(y)$\ which
has a damped oscillating behavior at infinity up to a constant, is 
\begin{equation}
Y(y)=\frac{(-i)\mathcal{W}_{M}(-ik\mathbf{n},1/2,2iky)}{2ky}  \label{Y2}
\end{equation}%
where $\mathcal{W}_{M}$ \ is the Whittaker function. The final general
solution will be a superposition of all possible solutions in the form 
\begin{equation}
H_{TN\text{ }\otimes \text{ }AH}(y,r)=1+Q_{M2}\int_{0}^{\infty }dkq(k)\frac{%
(-i)\mathcal{W}_{M}(-ik\mathbf{n},1/2,2iky)}{2ky}R_{k}(r)  \label{HTNAH}
\end{equation}%
where as before $q(k)$\ can be determined by looking at some near
horizon/bolt limit. For $r\simeq \pi n$\ and $y<<\mathbf{n}$, \ the
transverse metric reduces to $\mathbb{R}^{4}\otimes \mathbb{R}^{2}\otimes
S^{2}$\ since the Taub-NUT metric (\ref{dstn4hashi}) reduces to $\mathbb{R}%
^{4}$\ with line element $d\mathfrak{z}^{2}+\mathfrak{z}^{2}d\Omega _{3}^{2}$%
, where $\mathfrak{z}=2\sqrt{2\mathbf{n}y}.$\ The transverse radial distance
to the bolt is given by $\mathcal{R}=\sqrt{\mathfrak{z}^{2}+\widetilde{r}^{2}%
}$. Comparing the metric function (\ref{HTNAH}) with that of an M2-brane in
transverse flat space, we can fix the measure function to be $q(k)\propto
k^{5}$. Absorbing the constant into the M2-brane charge, we obtain%
\begin{equation}
H_{TN\text{ }\otimes \text{ }AH}(y,r)=1-iQ_{M2}\int_{0}^{\infty }dkk^{4}%
\frac{\mathcal{W}_{M}(-ik\mathbf{n},1/2,2iky)}{y}R_{k}(r)  \label{HfinTNAH}
\end{equation}%
as the metric function of M2-brane solution (\ref{ds11m2withTN}). Note that
the Whittaker function in the integrand is pure imaginary, yielding a
real-valued $H_{TN\text{ }\otimes \text{ }AH}(y,r)$.

Compactifying over the circle parametrized by $\sigma $\ (and noting that $%
\partial /\partial \sigma $\ is a Killing vector) we find the NSNS fields%
\begin{equation}
\begin{array}{rcl}
\Phi & = & \frac{3}{4}\ln \left( \frac{H^{1/3}}{f}\right) \\ 
B_{\mu \nu } & = & 0%
\end{array}
\label{tnXAHNSNS}
\end{equation}%
\ and RR fields

\begin{equation}
\begin{array}{rcl}
C_{\beta } & = & 2\mathbf{n}\cos (\alpha ) \\ 
A_{tx_{1}x_{2}} & = & H^{-1}.%
\end{array}
\label{tnXAHRR}
\end{equation}%
The metric in ten dimensions will be given by 
\begin{eqnarray}
ds_{10}^{2} &=&H_{TN\text{ }\otimes \text{ }AH}(y,r)^{-1/2}f^{-1/2}\left(
-dt^{2}+dx_{1}^{2}+dx_{2}^{2}\right) +H_{TN\text{ }\otimes \text{ }%
AH}(y,r)^{1/2}f^{-1/2}ds_{AH}^{2}+  \notag \\
&+&H_{TN\text{ }\otimes \text{ }AH}(y,r)^{1/2}f^{1/2}\left(
dy^{2}+y^{2}(d\alpha ^{2}+\sin ^{2}(\alpha )d\beta ^{2})\right) .
\label{ds10tnXAH}
\end{eqnarray}%
This represents a D2$\perp $D6(2) system that preserves 1/4 of \ the
supersymmetry. We have explicitly checked that the above 10-dimensional
metric, with the given dilaton, one and three forms, is a solution to the
10-dimensional supergravity equations of motion.

\begin{figure}[tbp]
\centering                                                  
\begin{minipage}[c]{.3\textwidth}
        \centering
        \includegraphics[width=\textwidth]{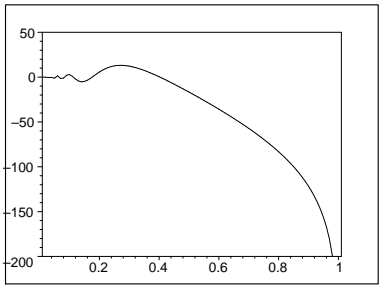}
    \end{minipage}
\caption{{} Numerical solution of equation (\ref{Y0fyforEHAH}) for $%
Y_{k}/10^{5}$ as a function of $\frac{1}{y}$ for non-zero separation
constant $k$. The Eguchi-Hanson parameter $a$ is set to one and so for $%
y\approx a,$ the function $Y_{k}$ diverges and for $y\approx \infty $, it
vanishes.}
\label{fig3}
\end{figure}

\bigskip The third possible M2-brane solution is given by%
\begin{equation}
ds_{11}^{2}=H(y,r)^{-2/3}\left( -dt^{2}+dx_{1}^{2}+dx_{2}^{2}\right)
+H(y,r)^{1/3}\left( ds_{EH}^{2}+ds_{AH}^{2}\right)  \label{ds11m2withEH}
\end{equation}%
where 
\begin{eqnarray}
ds_{EH}^{2} &=&\frac{y^{2}}{4g(y)}\left[ d\sigma +\cos (\alpha )d\beta %
\right] ^{2}+g(y)dy^{2}+\frac{y^{2}}{4}\left( d\alpha ^{2}+\sin ^{2}(\alpha
)d\beta ^{2}\right)  \label{dseh4} \\
g(r) &=&\left( 1-\frac{a^{4}}{y^{4}}\right) ^{-1}.
\end{eqnarray}%
is the Eguchi-Hanson metric $ds_{EH}^{2}$. \ In this case, after separation
of variables by the relation (\ref{Hyrsep}), we find the same differential
equation for $R(r)$\ as given in equation (\ref{M2eqR}), and the
differential equation for $Y(y)$ is 
\begin{equation}
y(y^{4}-a^{4})Y_{k}^{\prime \prime }(y)+(3y^{4}+a^{4})Y_{k}^{\prime
}(y)+k^{2}y^{5}Y_{k}(y)=0.  \label{Y0fyforEHAH}
\end{equation}%
While an analytic closed solution for the differential equation (\ref%
{Y0fyforEHAH}) is not available, the numerical solution shows that it has a
damped oscillating behaviour at infinity which diverges at $y\simeq a.$\ A
typical numerical solution of \ $Y_{k}(y)$\ is given in figure \ref{fig3}. 
The general solution will be a superposition of all possible solutions in
the form%
\begin{equation}
H_{EH\text{ }\otimes \text{ }AH}(y,r)=1+Q_{M2}\int_{0}^{\infty
}dkr(k)Y_{k}(y)R_{k}(r)  \label{HEHAH0}
\end{equation}%
and we determine $r(k)$\ by looking at a near horizon/bolt limit. At $%
r\simeq \pi n$\ and $y\simeq a$, \ the transverse metric reduces to $\mathbb{%
R}^{2}\otimes S^{2}\otimes \mathbb{R}^{2}\otimes S^{2}$, since the
Eguchi-Hanson metric (\ref{dseh4}) for $y=a(1+\epsilon ^{2})$, where $%
\epsilon \ll 1$, reduces to 
\begin{equation}
ds_{EH}^{2}\approx \mathbf{z}^{2}d\sigma ^{2}+d\mathbf{z}^{2}+\frac{a^{2}}{4}%
\left( d\alpha ^{2}+\sin ^{2}(\alpha )d\beta ^{2}\right)  \label{EHatyeqa}
\end{equation}%
which is the metric of $\mathbb{R}^{2}\otimes S^{2}$\ and $\mathbf{z}%
=a\epsilon .$\ In this limit and for small $a$, the differential equation (%
\ref{Y0fyforEHAH}) has the real solution 
\begin{equation}
Y(\widehat{y})=-i\frac{I_{1}(ik\widehat{y})}{\widehat{y}}  \label{YEHapp}
\end{equation}%
that vanishes at infinity where $\widehat{y}=a\epsilon ^{2}$. The transverse
radial distance to bolt is given by $\mathcal{R}=\sqrt{\mathbf{z}^{2}+\frac{%
a^{2}}{4}+\widetilde{r}^{2}}$. Comparing the metric function (\ref{HEHAH0})
with that of an M2-brane in transverse flat space, we fix the measure
function $r(k)\propto k^{5}$, yielding 
\begin{equation}
H_{EH\text{ }\otimes \text{ }AH}(y,r)=1+Q_{M2}\int_{0}^{\infty
}dkk^{5}Y_{k}(y)R_{k}(r)  \label{HfinEHAH}
\end{equation}%
for the metric function of the M2-brane solution (\ref{ds11m2withEH}), where
we have absorbed a constant into the M2-brane charge.

In this case, by compactification along the $\sigma $ direction of
Eguchi-Hanson metric, we find the NSNS fields%
\begin{equation}
\begin{array}{c}
\Phi =\frac{3}{4}\ln \left\{ \frac{H^{1/3}w^{2}}{4g}\right\} \\ 
B_{\mu \nu }=0%
\end{array}
\label{NSNSEH}
\end{equation}%
and RR fields%
\begin{equation}
\begin{array}{c}
C_{\beta }=a\cos (\alpha ) \\ 
A_{tx_{1}x_{2}}=\frac{1}{H}%
\end{array}
\label{RREH}
\end{equation}%
and metric 
\begin{eqnarray}
ds_{10}^{2} &=&\frac{w}{2}\{H^{-1/2}g^{-1/2}\left(
-dt^{2}+dx_{1}^{2}+dx_{2}^{2}\right) +  \notag \\
&+&H^{1/2}g^{-1/2}ds_{AH}^{2}+H^{1/2}g^{1/2}a^{2}\{dw^{2}+\frac{w^{2}}{4g}%
\left( d\alpha ^{2}+\sin ^{2}\alpha d\beta ^{2}\right) \}\}  \label{ds10AHEH}
\end{eqnarray}%
where $w=\frac{y}{a}.$\ The metric describes an intersecting D2/D6 system
where D2 is localized along the world-volume of the D6-brane and the
world-volume of the D6 brane transverse to D2 is just Atiyah-Hitchin space.
We note that in the large $w$\ limit, the metric (\ref{ds10AHEH}), reduces
to the metric%
\begin{equation}
ds_{10}^{2}=\frac{w}{2}%
\{-dt^{2}+dx_{1}^{2}+dx_{2}^{2}+ds_{AH}^{2}+a^{2}(dw^{2}+\frac{w^{2}}{4}%
\left( d\alpha ^{2}+\sin ^{2}\alpha d\beta ^{2}\right) )\}
\end{equation}%
which is again a 10D locally asymptotically flat metric with Kretchmann
invariant 
\begin{equation}
R_{\mu \nu \rho \sigma }R^{\mu \nu \rho \sigma }=\frac{A(a,r)}{w^{2}}+\frac{%
224}{w^{6}a^{4}}  \label{K2}
\end{equation}%
which vanishes at large $w.$\ $A(a,r)$\ is a complicated function of the
Eguchi-Hanson parameter $a$\ and Atiyah-Hitchin metric functions $a(r),b(r)$%
\ and $c(r).$\ All the components of the Riemann tensor in the orthonormal
basis approach zero at $w\rightarrow \infty $.

Since $\partial /\partial \psi $\ is a Killing vector we can further reduce
the metrics (\ref{ds10tnXAH}) and (\ref{ds10AHEH}) along the $\psi $\
direction of the Atiyah-Hitchin space at large $r$\ where $a(r)=b(r)$\ (up
to terms of order of $e^{-r/n}$). However the result of this
compactification is not the same as the reduction of the M-theory solution
over a torus, which is compactified type-IIB theory. To get the compactified
type-IIB theory, we must T-dualize the metrics (\ref{ds10tnXAH}) and (\ref%
{ds10AHEH}) first and then compactify the resultant type-IIB solutions along
the $\psi $\ direction of the Atiyah-Hitchin space.

Finally, the fourth possible M2-brane solution is given by%
\begin{equation}
ds_{11}^{2}=H(r_{1},r_{2})^{-2/3}\left( -dt^{2}+dx_{1}^{2}+dx_{2}^{2}\right)
+H(r_{1},r_{2})^{1/3}\left( ds_{AH_{1}}^{2}+ds_{AH_{2}}^{2}\right)
\label{AHAH}
\end{equation}%
where $ds_{AH_{1}}^{2}$\ and $ds_{AH_{2}}^{2}$\ are given by two copies of (%
\ref{AHmetric}) with coordinate systems $(r_{1},\theta _{1},\phi _{1},\psi
_{1})$\ and $(r_{2},\theta _{2},\phi _{2},\psi _{2})$, respectively. \ In
this case, after separation of variables by the relation 
\begin{equation}
H(r_{1},r_{2})=1+Q_{M2}\widetilde{R}(r_{1})R(r_{2})  \label{sepofvar}
\end{equation}%
\ we find two differential equations%
\begin{equation}
a_{1}c_{1}r_{1}^{2}\frac{d^{2}\widetilde{R}_{k}(r_{1})}{dr_{1}^{2}}%
+\{a_{1}c_{1}r_{1}-\frac{1}{2}%
r_{1}[(a_{1}-b_{1})^{2}+(b_{1}-c_{1})^{2}-a_{1}^{2}-c_{1}^{2}]\}\frac{d%
\widetilde{R}_{k}(r_{1})}{dr_{1}}+k^{2}a_{1}b_{1}^{2}c_{1}\widetilde{R}%
_{k}(r_{1})=0  \label{R1}
\end{equation}%
\ 
\begin{equation}
a_{2}c_{2}r_{2}^{2}\frac{d^{2}R_{k}(r_{2})}{dr_{2}^{2}}+\{a_{2}c_{2}r_{2}-%
\frac{1}{2}r_{2}[(a_{2}-b_{2})^{2}+(b_{2}-c_{2})^{2}-a_{2}^{2}-c_{2}^{2}]\}%
\frac{dR_{k}(r_{2})}{dr_{2}}-k^{2}a_{2}b_{2}^{2}c_{2}R_{k}(r_{2})=0.
\label{R2}
\end{equation}%
We note the differential equation for $R_{k}(r_{2})$\ is the same as
equation (\ref{M2eqR}), that its typical solution is presented in figures %
\ref{M2Rnumeric} and \ref{M2Ratlarger}\ for near bolt and near infinity
regions. As before, an analytic closed solution for the first differential
equation (\ref{R1}) is not available. However the numerical solution
(presented in figures \ref{figRsc}\ and \ref{M2Ratlargersc} in section \ref%
{sec:sc}) shows that it has diverging behaviour at $r_{1}\simeq \pi n_{1}$\
and damped oscillating behaviour at infinity. In fact we note later that for 
$r_{1}\simeq \pi n_{1}$\ (the only reliable region in figure \ref{figRsc}),
the function $\widetilde{R}_{k}(r_{1})$\ has a logarithmic divergence.

The general solution will be a superposition of all possible solutions in
the form%
\begin{equation}
H_{AH\text{ }\otimes \text{ }AH}(r_{1},r_{2})=1+Q_{M2}\int_{0}^{\infty
}dks(k)\widetilde{R}_{k}(r_{1})R_{k}(r_{2})  \label{Hah2}
\end{equation}%
where $s(k)\propto k^{5}$\ following previous near-horizon/bolt arguments
and comparing the metric function (\ref{Hah2}) with that of M2-brane in
transverse flat space . In this case as $r_{1}\simeq \pi n_{1}$\ and $%
r_{2}\simeq \pi n_{2}$, \ the transverse metric in (\ref{AHAH}) reduces to $%
\mathbb{R}^{4}\otimes \mathbb{R}^{4}$\ with line element $d\mathfrak{z}%
_{1}^{2}+d\mathfrak{z}_{2}^{2}+\mathfrak{z}_{1}^{2}d\Omega _{3}^{2}+%
\mathfrak{z}_{2}^{2}d\Omega _{3}^{^{\prime }2}$, where\textbf{\ }$\mathfrak{z%
}_{i}=2\sqrt{2n_{i}r_{i}}$\textbf{\ }and the transverse radial distance to
the bolt is given by $\mathcal{R}=\sqrt{\mathfrak{z}_{1}^{2}+\mathfrak{z}%
_{2}^{2}}$. We find%
\begin{equation}
H_{AH\text{ }\otimes \text{ }AH}(r_{1},r_{2})=1+Q_{M2}\int_{0}^{\infty
}dkk^{5}\widetilde{R}_{k}(r_{1})R_{k}(r_{2})  \label{HAHAH2}
\end{equation}%
where we have absorbed additional constants into the M2-brane charge.

Since $\partial /\partial \psi _{1}$\ and $\partial /\partial \psi _{2}$\
are not Killing vectors except for $a_{1}(r_{1})=b_{1}(r_{1})$\ and $%
a_{2}(r_{2})=b_{2}(r_{2})$, we cannot use the reduction relations (\ref%
{FKKreduced1}) and (\ref{FKKreduced2}) to find 10D type IIA fields
explicitly. However, we note that at large $r_{1}$\ (or at large $r_{2}$),
we have $a_{1}(r_{1})=b_{1}(r_{1})$\ (or $a_{2}(r_{2})=b_{2}(r_{2})$) up to
terms of order of $e^{-r_{1}/n_{1}}$\ (or $e^{-r_{2}/n_{2}}$). Consequently,
although we cannot explicitly write the 10D brane system fields
corresponding to the M2-brane solution (\ref{ds11m2}), the reduction
relations (\ref{FKKreduced1}) and (\ref{FKKreduced2}) at large $r_{1}$\ (or
at large $r_{2}$) yield the explicit fields in 10D that describe a D2$\perp $%
D6(2) system that preserves 1/4 of the supersymmetry \cite{CGMM2}.

To summarize, all M2-brane solutions with an Atiyah-Hitchin space in the
transverse geometry preserve 1/4 of the supersymmetry even though
Atiyah-Hitchin space has a bolt-like fixed point at $r=\pi n.$\ This
behaviour is completely different from what was observed in\textbf{\ }\cite%
{CGMM2}, where the only M2-brane solutions preserving any supersymmetry had
NUT-like fixed points (i.e. of less than maximal co-dimensionality). \
M2-brane solutions with transverse Taub-Bolt spaces of various
dimensionalities were not supersymmetric. Unlike these cases, the
four-dimensional Atiyah-Hitchin space is self-dual hyper-K\"{a}hler, thereby
preserving some supersymmetry in the associated M2-brane solutions.

\section{A Second set of M2-brane solutions}

\label{sec:sc}{\Large \ \ \ \ \ }

A different set of M2-brane solutions can be obtained by reversing the sign
of the separation constant $k^{2}$ in the separated differential equations
for $Y(y)$ and $R(r).$ As an example, by taking $k\rightarrow i\widetilde{k}$
in the separable equations of Atiyah-Hitchin case, we find the solution of
the differential equation for $\widetilde{Y}(y)$ as 
\begin{figure}[tbp]
\centering                                    
\begin{minipage}[c]{.3\textwidth}
        \centering
        \includegraphics[width=\textwidth]{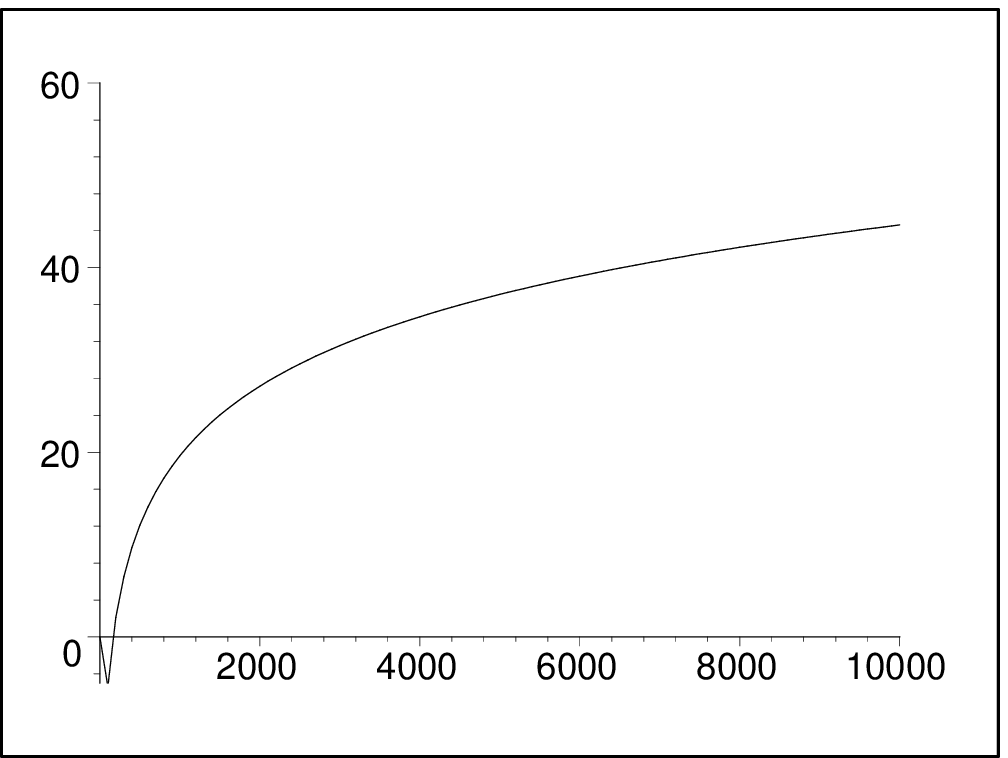}
    \end{minipage}
\caption{ Numerical solution of the radial equation (\ref{M2eqRsc}) for $%
\widetilde{R}_{\widetilde{k}}/10^{6}$ as a function of $\frac{1}{r-\protect%
\pi }$, where we set $n=1$ for simplicity. So for $r\approx \protect\pi $, $%
\widetilde{R}$ diverges logarithmically by $\ln (\frac{1}{r-\protect\pi })$.
Note that the plot is reliable only for large $\frac{1}{r-\protect\pi }$ .}
\label{figRsc}
\end{figure}
\begin{equation}
\widetilde{Y}(y)=\frac{K_{1}(\widetilde{k}y)}{y}  \label{Y1sc}
\end{equation}%
where $K_{1}$\ is the modified Bessel function, diverging at $y=0$ and
vanishing at infinity. The differential equation for $\widetilde{R}(r)$ is
given by

\begin{equation}
acr^{2}\frac{d^{2}\widetilde{R}_{\widetilde{k}}(r)}{dr^{2}}+\{acr-\frac{1}{2}%
r[(a-b)^{2}+(b-c)^{2}-a^{2}-c^{2}]\}\frac{d\widetilde{R}_{\widetilde{k}}(r)}{%
dr}+\widetilde{k}^{2}ab^{2}c\widetilde{R}_{\widetilde{k}}(r)=0.
\label{M2eqRsc}
\end{equation}%
Although the above equation does not have any analytic closed solution, we
can solve it numerically. Typical solutions are presented in figures \ref%
{figRsc}\ and \ref{M2Ratlargersc} for $r\simeq \pi n$ and large $r$ regions
respectively, where we set $n=1$. We note that for $r\simeq \pi n$ (the only
reliable region in figure \ref{figRsc}), the function $\widetilde{R}_{%
\widetilde{k}}(r)$ also has a logarithmic divergence given by 
\begin{equation}
\widetilde{R}_{\widetilde{k}}(r)\simeq -Y_{0}(\widetilde{k}(r-\pi n))=-\frac{%
2}{\pi }\{\ln (\frac{\widetilde{k}}{2})+\gamma +\ln (r-\pi n)\}+O\{(r-\pi
n)^{2}\}
\end{equation}%
where $Y_{0}$ is the Bessel function of the second kind.

Figure \ref{M2Ratlargersc} shows the behaviour of the radial function $%
\widetilde{R}_{\widetilde{k}}(r)$ at large $r$, given by%
\begin{equation}
\widetilde{R}_{\widetilde{k}}(r)\simeq \frac{\cos (\widetilde{k}r)}{r}
\end{equation}%
which is obtained by using the relations (\ref{abcatinfinity}). The final
general solution will be a superposition of all possible solutions and has
the form 
\begin{equation}
\widetilde{H}(y,r)=1+Q_{M2}\int_{0}^{\infty }d\widetilde{k}\widetilde{p}(%
\widetilde{k})\frac{K_{1}(\widetilde{k}y)}{y}\widetilde{R}_{\widetilde{k}}(r)
\label{scAH}
\end{equation}%
where $\widetilde{p}(\widetilde{k})$\ can be computed by comparing the
relation (\ref{scAH}) to that of a metric function of an M2-brane in a
transverse flat metric\ \ $\mathbb{R}^{4}\otimes \mathbb{R}^{2}\otimes S^{2}$%
, obtained by looking at the near bolt limit. We obtain%
\begin{equation}
\widetilde{H}_{AH}(y,r)=1+Q_{M2}\int_{0}^{\infty }d\widetilde{k}\widetilde{k}%
^{4}\frac{K_{1}(\widetilde{k}y)}{y}\widetilde{R}_{\widetilde{k}}(r)
\label{HAHss2}
\end{equation}%
as the second M2-brane solution (\ref{ds11m2}), absorbing a possible
constant into the charge\textbf{\ }$Q_{M2}$.

The other two alternative solutions for the Taub-NUT $\otimes $
Atiyah-Hitchin and Eguchi-Hanson $\otimes $ Atiyah-Hitchin could be derived
easily similar to the above case and so we do not present them here. In the
Atiyah-Hitchin $\otimes $\ Atiyah-Hitchin case with metric function (\ref%
{HAHAH2}), the transformation $k\rightarrow i\widetilde{k}$\ in (\ref{R1})
and (\ref{R2}), merely interchanges $\widetilde{R}_{k}(r_{1})\rightarrow R_{%
\widetilde{k}}(r_{1})$\ and $R_{k}(r_{2})\rightarrow \widetilde{R}_{%
\widetilde{k}}(r_{2})$ and so yields no new solution. 
\begin{figure}[tbp]
\centering                                    
\begin{minipage}[c]{.3\textwidth}
        \centering
        \includegraphics[width=\textwidth]{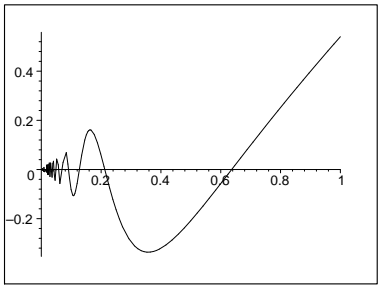}
    \end{minipage}
\caption{ Behavior of the radial function $\widetilde{R}_{\widetilde{k}}(r)$
as a function of $\frac{1}{r},$ near infinity.}
\label{M2Ratlargersc}
\end{figure}

\section{Embedding Atiyah-Hitchin space in an M5-brane metric}

\label{sec:M5}

The eleven dimensional M5-brane metric with an embedded Atiyah-Hitchin
metric has the following form%
\begin{equation}
\begin{array}{c}
ds_{11}^{2}=H(y,r)^{-1/3}\left(
-dt^{2}+dx_{1}^{2}+dx_{2}^{2}+dx_{3}^{2}+dx_{4}^{2}+dx_{5}^{2}\right) + \\ 
+H(y,r)^{2/3}\left( dy^{2}+ds_{AH}^{2}\right)%
\end{array}
\label{ds11m5p}
\end{equation}%
with field strength components%
\begin{equation}
\begin{array}{c}
F_{\psi \theta \phi y}=\frac{\alpha }{2}\sin (\theta )rac\frac{\partial H}{%
\partial r} \\ 
F_{\psi \theta \phi r}=-\frac{\alpha }{2r}\sin (\theta )ab^{2}c\frac{%
\partial H}{\partial y}.%
\end{array}
\label{FScompo}
\end{equation}%
We consider the M5-brane which corresponds to $\alpha =+1$;\ the $\alpha =-1$%
\ case corresponds to an anti-M5 brane.

The metric (\ref{ds11m5p}) is a solution to the eleven dimensional
supergravity equations provided $H\left( y,r\right) $ is a solution to the
differential equation 
\begin{equation}
r^{2}\frac{\partial ^{2}H}{\partial r^{2}}+r\{1+r(\frac{a^{\prime }}{a}+%
\frac{c^{\prime }}{c})\}\frac{\partial H}{\partial r}+b^{2}\frac{\partial
^{2}H}{\partial y^{2}}=0.  \label{HeqforM5}
\end{equation}%
This equation is straightforwardly separable. Substituting 
\begin{equation}
H(y,r)=1+Q_{M5}Y(y)R(r)
\end{equation}%
where $Q_{M5}$ is the charge on the M5-brane. The solution of the
differential equation for $Y(y)$ is 
\begin{equation}
Y(y)=\cos (ky+\varsigma )
\end{equation}%
and the differential equation for $R(r)$ is given by the equation (\ref%
{M2eqR}). The numerical solution of this equation near $r\simeq \pi n,$ is
again given by figure \ref{M2Rnumeric}; the final solution is a
superposition of all possible solutions 
\begin{equation}
H(y,r)=1+Q_{M5}\int_{0}^{\infty }dkh(k)\cos (ky)R_{k}(r)  \label{M5}
\end{equation}%
where $h(k)$ is the measure function.

To fix the measure function $h(k)$\ we compare the relation (\ref{M5}) to
that of a metric function of M5-brane in transverse flat metric\ \ $\mathbb{R%
}^{4}\otimes \mathbb{R}^{2}\otimes S^{2}$, obtained by looking at the near
bolt limit $r\simeq \pi n.$\ In this case the radial function in (\ref{M5})
reduces to (\ref{RatreqPi}). Hence for reduction of the metric function (\ref%
{M5}) to $1+\frac{Q_{M5}}{\mathcal{R}^{3}}$\ where $\mathcal{R}=\sqrt{y^{2}+%
\widetilde{r}^{2}}$\ and $\widetilde{r}=r-\pi n$, we must fix the measure
function to be $h(k)\propto k^{2}$, giving%
\begin{equation}
H_{AH}(y,r)=1+Q_{M5}\int_{0}^{\infty }dkk^{2}\cos (ky)R_{k}(r)
\label{AHgeneralHforM5}
\end{equation}%
as the metric function of M5-brane solution (\ref{ds11m5p}), where we absorb
the constant into the M5-brane charge. We note that in equation (\ref%
{AHgeneralHforM5}), $R_{k}(r)$ for $r\simeq \pi n$ approaches the numerical
solution, presented in figure \ref{M2Rnumeric} and for large $r,$ is given
by the limit of the radial function in (\ref{M2eqR}) or equivalently by (\ref%
{Ratlarger}).

As with the Atiyah-Hitchin-based M2 solution we can dimensionally reduce our
M5 solution to find 10D type IIA fields at large $r$\ , since $\partial
/\partial \psi $\ is a Killing vector and $a(r)=b(r)$\ up to terms of order
of $e^{-r/n}$. The resulting fields describe an NS5$\perp $D6(5) system
which preserves 1/4 of the supersymmetry. We expect in the decoupling limit
of our solution that is in the limit of vanishing string coupling, the
theory on the worldvolume of the NS5-branes is a type IIA little string
theory \cite{CGMM5}.

A different M5-brane solution can be obtained by reversing the sign of the
separation constant $k^{2}$\ in the separated differential equations
obtained from (\ref{HeqforM5}).\ In this case, by taking $k\rightarrow i%
\widetilde{k}$\ , we find another solution in the form of%
\begin{equation}
\widetilde{H}_{AH}(y,r)=1+Q_{M5}\int_{0}^{\infty }d\widetilde{k}\widetilde{k}%
^{2}e^{-\widetilde{k}y}\widetilde{R}_{\widetilde{k}}(r)
\label{AHgeneralHforM5ss}
\end{equation}%
where numerical plot of the function $\widetilde{R}_{\widetilde{k}}$\ is
given in figures \ref{figRsc}\ and \ref{M2Ratlargersc} for $r\simeq n\pi $\
and large $r$\ regions, respectively. \ Although this is formally a
solution, the integral in (\ref{AHgeneralHforM5ss}) is not convergent for
all values of $y$. To make the integral convergent for $y<0$, one can
replace $e^{-\widetilde{k}y}$\ by $e^{-\widetilde{k}\left| y\right| }$, but
only at the price of introducing a source term at $y=0$\ in the
corresponding Laplace equation for $\ \widetilde{H}_{AH}(y,r).$

As before, we do not have a representation of 10D fields since \ $\partial
/\partial \psi $\ is not a Killing vector except for $a(r)=b(r).$\ However,
we note that at large $r$, we have $a(r)=b(r)$\ up to terms of order of $%
e^{-r/n}$\ and in this case, by using the reduction relations (\ref%
{FKKreduced1}) and (\ref{FKKreduced2}), the explicit fields in 10D describe
an NS5$\perp $D6(5) system that preserves 1/4 of the supersymmetry \cite%
{CGMM5}. So although the 10D brane system fields corresponding to M5-brane
solution (\ref{ds11m5p}) with metric functions (\ref{AHgeneralHforM5}) or (%
\ref{AHgeneralHforM5ss}) cannot be written explicitly, the asymptotic
behavior of the system is given by the known fields of the\ NS5$\perp $D6(5)
system that preserves 1/4 of the supersymmetry. As the last case, we expect
in the decoupling limit of solution corresponding to (\ref{AHgeneralHforM5ss}%
), the theory on the worldvolume of the NS5-branes is a type IIA little
string theory \cite{CGMM5}.

We see again that the M5-brane solution (\ref{AHgeneralHforM5}) with an
Atiyah-Hitchin space in the transverse geometry preserves 1/4 of the
supersymmetry even though the Atiyah-Hitchin space has a bolt-like fixed
point at $r=n\pi .$\ This behaviour is completely different from that found
in \cite{CGMM5}, where the only M5-brane solutions preserving any
supersymmetry had NUT-like fixed points (i.e. of less than maximal
co-dimensionality). \ The M5-brane solutions with four dimensional
transverse Taub-Bolt space were not supersymmetric. Unlike these cases, the
four-dimensional Atiyah-Hitchin space is self-dual hyper-K\"{a}hler, thereby
preserving some supersymmetry in the associated M5-brane solutions.

\section{Conclusion}

By embedding Atiyah-Hitchin space into M-theory, we have found new classes
of \ 2-brane and 5-brane solutions to $D=11$\ supergravity. These exact
solutions are new M2- and M5-brane metrics with metric functions (\ref%
{AHgeneralH}), (\ref{HfinTNAH}), (\ref{HfinEHAH}), (\ref{HAHAH2}), (\ref%
{HAHss2}), (\ref{AHgeneralHforM5}) and (\ref{AHgeneralHforM5ss}) -- these
are the main results of this paper. The common feature of both solutions is
that the brane function is a convolution of an exponentially decaying
`radial' function (for both branes) with a damped oscillating one. The
`radial' function vanishes far from the branes and diverges logarithmically
near the brane core. \ The same logarithmic divergence near the brane
happens in embedding of \ Eguchi-Hanson metric in M-theory where the
divergence is milder than $\frac{1}{r},$\ as in the case of embedding
Taub-NUT space. Indeed, all of these properties of our solutions are similar
to those previously obtained \cite{CGMM2,CGMM5} for the embedding of
Eguchi-Hanson and Taub-NUT spaces.

However our solutions have a feature that is quite distinct from these
predecessors: they are bolt solutions (i.e. solutions whose fixed points in
the transverse space have maximal dimensionality) that preserve 1/4 of the
supersymmetry due to the self-dual hyper-K\"{a}hler character of the
Atiyah-Hitchin metric. This is in contrast to earlier brane solutions of
this type \cite{CGMM2,CGMM5}, for which supersymmetry could only be
preserved for NUT-like transverse metrics; their bolt counterparts did not
preserve any supersymmetry.

Dimensional reduction of the M2 solutions to ten dimensions gives us
intersecting IIA D2/D6 \ configurations that preserve 1/4 of the
supersymmetry. For the M5 solutions, dimensional reduction yields IIA\
NS5/D6 brane systems overlapping in five directions.

In the standard case, the system of N$_{5}$\ NS5-branes located at N$_{6}$\
D6-branes can be obtained by dimensional reduction of \ N$_{5}$N$_{6}$\
coinciding images of M5-branes in the flat transverse geometry. In this
case, the worldvolume theory (the little string theory) of the IIA
NS5-branes, in the absence of D6-branes, is a non-local non-gravitational
six dimensional theory \cite{seiberg}. This theory has (2,0) supersymmetry
(four supercharges in the 4\ representation of Lorentz symmetry $Spin(5,1)$)
and an R-symmetry $Spin(4)$\ remnant of the original ten dimensional Lorentz
symmetry. The presence of the D6-branes breaks the supersymmetry down to
(1,0), with eight supersymmetries. Since we found that our solutions
preserve 1/4 of the supersymmetry, we expect that the theory on NS5-branes
is a new little string theory. \ 

We note that in the limit of large $r$, where $a(r)=b(r)$, the decoupling
limits of M2 and M5 Atiyah-Hitchin based brane solutions are qualitatively
the same as what was found in references [6, 7, 8] for corresponding
Taub-NUT based solutions. Although the asymptotic holographic duals of the
decoupled theories are known, it would therefore be interesting to find the
complete structures of the holographic duals of the decoupled theories. We
leave this for future study.

We note that applying T-duality on our M2 and M5 solutions generates new IIB
configurations. For example, by T-dualization along any spatial directions
of the NS5 brane, type IIA NS5$\perp $D6(5) system changes to type IIB NS5$%
\perp $D5(4) brane configuration, overlapping in four directions.

The worldvolume theory of the IIB NS5-branes, in the absence of D5-branes,
is a little string theory with (1,1) supersymmetry. The presence of the
D5-brane, which has one transverse direction relative to NS5 worldvolume,
breaks the supersymmetry down to eight supersymmetries. This is in good
agreement with the number of supersymmetries in 10D IIB theory: T-duality
preserves the number of original IIA supersymmetries, that is eight.

Moreover we conclude that the new IIA and IIB little string theories are
T-dual: the actual six dimensional T-duality is the remnant of the original
10D T-duality after toroidal compactification.

\bigskip

{\Large Acknowledgments}

This work was supported by the Natural Sciences and Engineering Research
Council of Canada.

\end{document}